\documentclass[prl,12pt,preprint,a4paper,showkeys,showpacs]{revtex4}

\usepackage[dvips]{graphics}

\begin{document}

\title{On the entropy of classical systems with long-range interaction.}
\author{T.\ M.\ Rocha Filho, A.\ Figueiredo \& M.\ A.\ Amato}
\affiliation{Instituto de F\'\i{}sica, Universidade de
Bras\'\i{}lia\\ CP: 04455, 70919-970 - Bras\'\i{}lia, Brazil}

\begin{abstract}
We discuss the form of the entropy for classical hamiltonian systems with long-range
interaction using the Vlasov equation which describes the dynamics of a $N$-particle in the limit $N\rightarrow\infty$.
The stationary states
of the hamiltonian system are subject to infinite conserved quantities due to the Vlasov
dynamics. We show that the stationary states correspond to an extremum of the Boltzmann-Gibbs entropy,
and their stability is obtained from the condition that this extremum is a maximum.
As a consequence the entropy is a function of an infinite set of Lagrange multipliers that depend
on the initial condition. We also discuss
in this context the meaning of ensemble inequivalence and the temperature.
\end{abstract}


\maketitle

Systems interacting through long-range forces can present some types
of behavior that are not observed in more common systems. For example
inequivalence of the microcanonical and canonical ensembles,
negative specific heat, violent relaxation (rapid relaxation towards a non-gaussian quasi-stationary state),
superdiffusion and aging~\cite{r1,r2,r3,r3b,r3c}.
The most obvious example of long-range interaction is the gravitational force,
which is difficult to study due to its divergence at short distances.
A quite simple model that retains most of the behavior
found in realistic systems is the so-called Hamiltonian Mean Field (HMF) model
(see~\cite{r3d,r4,r5,r5b} and references therein).
Recently some authors proposed that
the Tsallis entropy could describe the statistical properties of such systems~\cite{r6,r7,r8,r8b}, although some
criticism has been raised in the literature~\cite{r5,r9,r10}. Here we present a different approach which
can shed some light on the problem, and also discuss some relevant aspects of the
meaning of temperature for long-range interacting systems.

Let us first consider a system of N identical particles described by the Hamiltonian
\begin{equation}
H=\sum_{i=1}^N \frac{p_i^2}{2m}+\sum_{i<j=1}^N\phi({\bf r}_i-{\bf r_j}),
\label{eq0}
\end{equation}
with ${\bf p}_i$  and ${\bf r}_i$ the momentum and position of the $i$-th particle respectively,
andi $\phi$ is the interaction potential. The force is 
long-ranged if the potential decays at long distances as $|{\bf r}_i-{\bf r_j}|^{-\alpha}$
with $\alpha<D$, with $D$ the spatial dimension.
In the limit $N\rightarrow\infty$ the inter-particle correlations are negligible, and the
system is described in the mean-field approximation by the Vlasov equation~\cite{r11,r12}:
\begin{equation}
\frac{\partial f}{\partial t}+{\bf v}\cdot\frac{\partial f}{\partial\bf r}
+{\bf F}\cdot\frac{\partial f}{\partial\bf v}=0,
\label{eq1}
\end{equation}
where $f$ is the one-particle mass distribution function in phase space and
$
{\bf F}({\bf r})=-{\partial U({\bf r})}/{\partial\bf r}
$
is the mean-field force.
The mean-field potential is given by
$
U({\bf r})=\int\phi({\bf r}-{\bf r}')f({\bf p}',{\bf r}',t)d^D{\bf p}'d^D{\bf r}'.
$
The Casimir functionals
$
C_s[f]=\int s(f({\bf p},{\bf r},t)) d^D{\bf p}\:d^D{\bf r}
$
are conserved by the Vlasov dynamics,
where $s(f)$ is an arbitrary function of $f$. This implies that the Vlasov equation (\ref{eq1}) admits an infinity of
stable stationary solutions. Any distribution which is an extremum (maximum or minimum) of a Casimir $C_s$ for a given
function $s(f)$ is a stable stationary solution of the Vlasov equation.

However real systems have always a finite number of particles, and corrections of order $1/N$ must be considered
to take into account collisional processes, that are important in the very long-time
regime and for systems with a number of particles not sufficiently large.
The kinetic equation in either case can be obtained in different ways
and we refer the reader to Reference~\cite{r12} for details. Nevertheless even for finite $N$ the stationary
state of the Vlasov equation describes, for a sufficiently long time, a quasi-stationary state of the real system
(up to ${\mathcal O}(1/N)$ effects).

The Boltzmann-Gibbs entropy is given by
$
S=-\int f_N\:\log f_N
\:d^D{\bf r}_1\cdots d^D{\bf r}_N\:d^D{\bf p}_1\cdots d^D{\bf p}_N,
$
where $f_N$ is the complete $N$-particle distribution function. In the mean-field limit
($N\rightarrow\infty$) the distribution $f_N$ factorizes as
$
f_N({\bf r}_1,\ldots,{\bf r}_N,{\bf p}_1,\ldots,{\bf p}_N,t) =
f({\bf p}_1,{\bf r}_1,t)\cdots f({\bf p}_N,{\bf r}_N,t),
$
and the entropy is thus written as
\begin{equation}
S=-N\int f({\bf r},{\bf p},t)\log f({\bf r},{\bf p},t)\:d^D{\bf r}\:d^D{\bf p}.
\label{eq6}
\end{equation}
The maximization of the entropy $S$ subject to the energy and normalization constraints:
\begin{equation}
{\cal H}=\int \frac{p^2}{2m} f({\bf r},{\bf p},t)\:d^D{\bf r}\:d^D{\bf p}
+\frac{1}{2}\int \phi({\bf r}-{\bf r}')f({\bf r},{\bf p},t)f({\bf r}^\prime,{\bf p}^\prime,t)
\:d^D{\bf r}\:d^D{\bf r}'\:d^D{\bf p}\:d^D{\bf p}'=E,
\label{eq7}
\end{equation}
\begin{equation}
{\cal I}=\int f({\bf r},{\bf p},t)\:d^D{\bf r}\:d^D{\bf p}=1,
\label{eq8}
\end{equation}
leads to the usual Maxwell-Boltzmann distribution. The stationary states of the Vlasov dynamics
are obtained by maximizing a Casimir $C_s$ with the
constrains in eqs.~(\ref{eq7}) and~(\ref{eq8}). This Casimir plays the role of a ``generalized entropy'',
and the sign of its second variation yields the stability condition for the stationary state.
This is the procedure adopted in Reference~\cite{r5} to study the stability of stationary states
of the HMF model. This model describes a system of $N$ interacting planar rotors with
hamiltonian~(\ref{eq0}), ${\bf r}_i=\theta_i$, $m=1$ and interacting potential
$
\phi({\bf r}_i-{\bf r_j})=v(\theta_i-\theta_j)\equiv1-\cos(\theta_i-\theta_j),
$
and such that $p_i$ is the angular momentum conjugate to the angle $\theta_i$.
We note that in this model the time evolution presents violent relaxation
into a quasi-stationary state with a life-time diverging with increasing $N$ and
ensemble inequivalence, and is therefore a prototypical model for the study of long-range
interacting systems~\cite{r2,r4}.

In order to undestrand the nature of the entropy in systems with long-range interactions,
we suppose that all microstates compatible with the given constraints are equally probable. This amounts
to maximize the Boltzmann-Gibbs entropy in the mean-field limit (eq.~\ref{eq6})
with the constraints shown in eqs.~(\ref{eq7}) and~(\ref{eq8}) and all
the Casimirs, and for all possible functions $s$.
Of course this is not an easy task since we have an infinite non-enumerable set of constraints.
Nevertheless we can simplify this problem by restricting ourselves to analytical distribution functions
(i.\ e.\ functions of class $C^\infty$), which is not a restriction for almost all physical situations.
Following this reasoning, the only
non-vanishing Lagrange multipliers are those associated to the energy and  normalization constrains as well
as those associated with Casimirs defined by an analytic function $s$.
It is therefore equivalent to consider
only the Casimirs of the form
$
C_n[f]=\int f^n \: d^D{\bf r}\:d^D{\bf p},
$ with $n=1,2,3,\ldots$
The extremum of $S$ under these Casimir and eqs.~(\ref{eq7}), (\ref{eq8}) constraints
is equivalent to the extremum of the functional
$
F=\frac{1}{N}S-\lambda_H {\cal H}-\lambda_I {\cal I}-\sum_{n=2}^\infty \lambda_n C_n,
$
where $\lambda_H$, $\lambda_I$ and $\lambda_n$ are Lagrange multipliers.
The $1/N$ factor is introduced for convenience and is equivalent to a rescaling of the Lagrange
multipliers. Computing the functional derivative of $F$ with respect to $f$ we have:
\begin{equation}
\frac{\delta F[f]}{\delta f}=-\log f-1-\lambda_I-\lambda_H e({\bf r},{\bf p})
-\sum_{n=2}^\infty n\lambda_n f^{n-1}=0,
\label{eq12}
\end{equation}
with
$
e({\bf r},{\bf p})={p^2}/{2m}+U({\bf r}),
$
and thus
\begin{equation}
G(f)\equiv\log f+\lambda_I+1+\sum_{n=1}^\infty n\lambda_n f^{n-1}=-\lambda_H e({\bf r},{\bf p}).
\label{eq14}
\end{equation}
Supposing the function $G$ is invertible we obtain
$
f({\bf r},{\bf p})=G^{-1}(-\lambda_H e({\bf r},{\bf p}))\equiv\Phi(e({\bf r},{\bf p})).
$
Distributions of this form are precisely
the stationary states of the Vlasov equation.
Since $e({\bf r},{\bf p})$ depends on $f({\bf r},{\bf p})$ the latter is given
self-consistently. For a homogeneous
system, the mean force vanishes and $e({\bf r},{\bf p})=p^2/2m$.
At this point we note that if $f$ never vanishes, then $G$
is analytic as a function of $f$, and
any analytic function $G$ can be written in the form of eq.~(\ref{eq14}), i.~e.\
it is obtained considering only the Casimirs with $s(f)=f^n$.
We conclude that we can take the Lagrange
multipliers associated with Casimirs other than $C_n[f]$ equal to zero.

The stability of a stationary state is usually determined by requiring that it is an
extremum (maximum or minimum) of a given Casimir under the energy and normalization constraints~\cite{r5}.
In the present approach, stability is associated to the condition of maximal entropy,
i.\ e.\ neither a minimum nor a saddle point. The equivalence of criteria should also be proved.
In this letter we will restrict ourselses to the case of homogeneous stationary states of
the HMF model for the sake of a more succinct presentation.

Similarly to Reference~\cite{r5}
we first define the $x$ and $y$ components of the total magnetization of the system of plane rotors by
\begin{equation}
M_x[f]=\int f(\theta,p,t) \cos\theta\:d\theta\:dp;\hspace{5mm}
M_y[f]=\int f(\theta,p,t) \sin\theta\:d\theta\:dp.
\label{eq17}
\end{equation}
The second variation of $F[f]$ computed at the extremum $f_0(\theta,p)$ must be negative for a maximum.
Thus
\begin{equation}
\int d\theta\:dp\:\left[-\frac{1}{2 f_0}
-\frac{1}{2}\sum_{n=2}^\infty n(n-1)\lambda_nf_0t^{n-2}\right]\delta f^2
+\frac{\lambda_H}{2}\left[\left(M_x[\delta f]\right)^2+\left(M_y[\delta f]\right)^2\right]<0.
\label{eq18}
\end{equation}
Differentiating eq.~(\ref{eq12}) with respect to $p$ and using the result in eq.~(\ref{eq18})
we have:
\begin{equation}
\lambda_H\int d\theta\:dp\:\frac{p}{f_0'}\:\delta f(\theta,p,t)^2
+\lambda_H\left[\left(M_x[\delta f]\right)^2+\left(M_y[\delta f]\right)^2\right]<0.
\label{eq20}
\end{equation}
Introducing the Fourier expansion
$
\delta f(\theta,p)=\sum_n\left[c_n(p)\cos n\theta+s_n(p)\sin n\theta\right],
$
into eq.~(\ref{eq20}) and using $\lambda_H>0$ we obtain
\begin{equation}
\int dp\:\frac{p}{f_0'}\left[2\pi c_0^2+\pi\sum_{n\geq 1}\left(c_n^2+s_n^2\right)\right]
+\left[\pi\int dp\:c_1\right]^2+\left[\pi\int dp\:s_1\right]^2<0.
\label{eq22}
\end{equation}
The type of stationary distribution function expected is an even function monotonously
increasing for $p<0$ and monotonously decreasing for $p>0$, which implies that $p/f_0'<0$.
Since the contribution of the terms with $c_0$, $c_n$ and $s_n$ with $n>1$ are always non-positive,
the maximum condition eq.~(\ref{eq22}) can be written as
$
W[c_1]+W[s_1]<0,
$
with
\begin{equation}
W[h]\equiv\int dp\:\pi\frac{p}{f_0'}h(p)^2+\pi^2\left[\int dp\:h(p)\right]^2.
\label{eq24}
\end{equation}
Using the Cauchy-Schwarz inequality we have that
\begin{equation}
\left[\int dp\: h(p)\right]^2=\left[\int dp\: \frac{h(p)\sqrt{-p/f_0'(p)}}{\sqrt{-p/f_0'(p)}}\right]^2
\leq\int dp\: \frac{f_0'(p)}{p}\times\int dp\:\frac{h(p)^2p}{f_0'(p)},
\label{eq25}
\end{equation}
which implies
$
W[h]\leq\pi\int dp\:\left[{p\:h(p)^2}/{f_0'(p)}\right]\cdot
\left[1+\pi\int dp\:{f_0'(p)}/{p}\right].
$
Since $p/f_0'(p)<0$ we finally obtain:
$
1+\pi\int dp\:{f_0'(p)}/{p}<0,
$
which coincides with the stability criterion obtained in Reference~\cite{r5}, up to a different normalization
for $f_0$. Thus the condition of maximal entropy subject to all Casimir invariants leads to the same
condition as obtained from the non-linear stability analysis of Yamaguchi et al.~\cite{r5}
by imposing that the state is the extremum of a given Casimir.

\begin{figure}
\begin{center}
\scalebox{0.33}{{\includegraphics{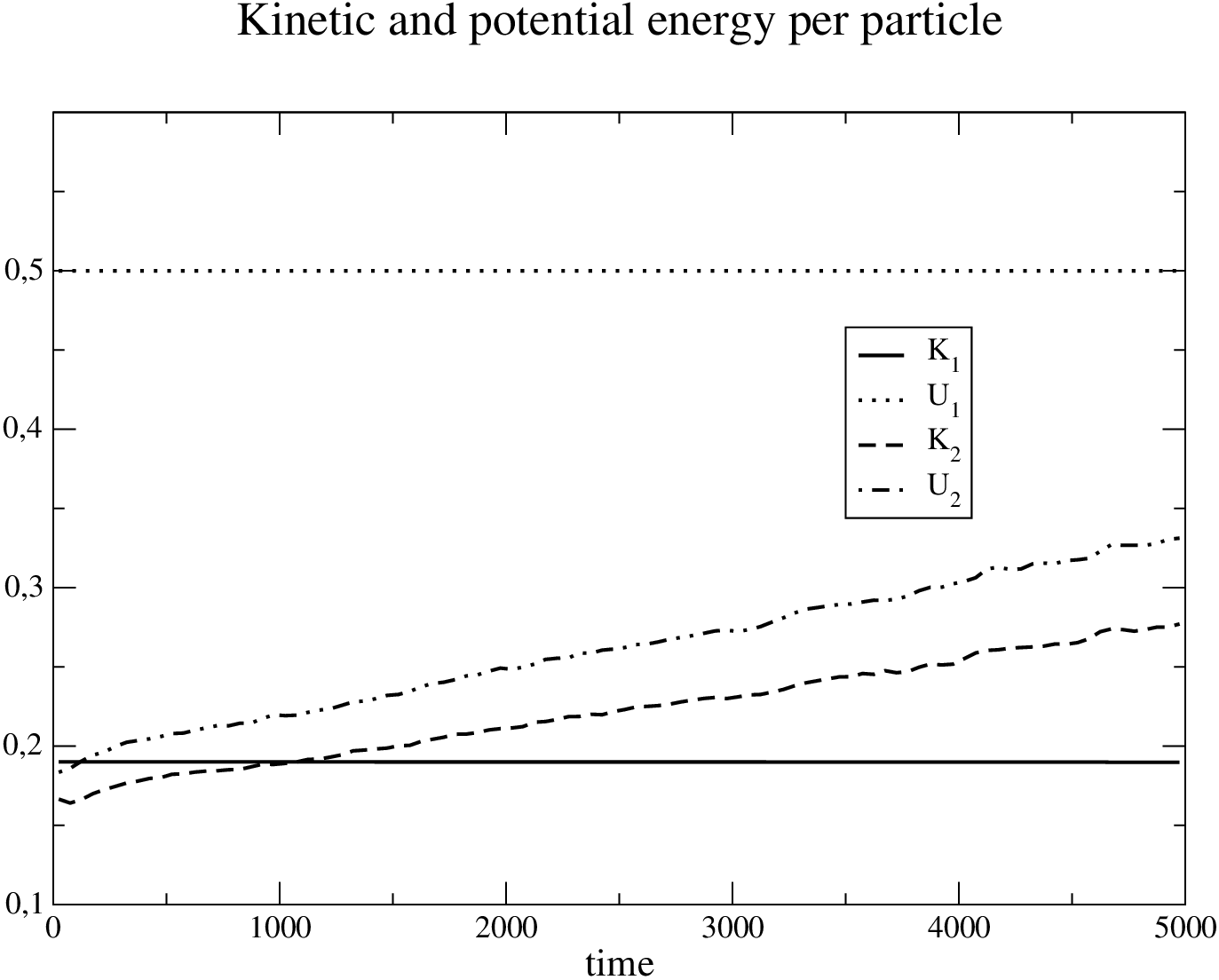}}}
\end{center}
\caption{Kinetic and potential energies per particle with $N_1=100,000$ (long-range) and
$N_2=100$ (short-range) obtained from $50$ simulations.
The initial energies per particle are $U_1=0.69$ and $U_2=0.35$, and
the time-step used is $\Delta t=0.2$ with an error $\Delta E/E\approx 10^{-8}$.
We let both system evolve without coupling until $t=100$ such that system 2 is in a gaussian velocity
distribution.}
\label{fig1}
\end{figure}

\begin{figure}
\hspace*{1mm}\noindent
\begin{minipage}{1in}
\vspace{0mm}
\raggedright
\scalebox{0.33}{{\includegraphics{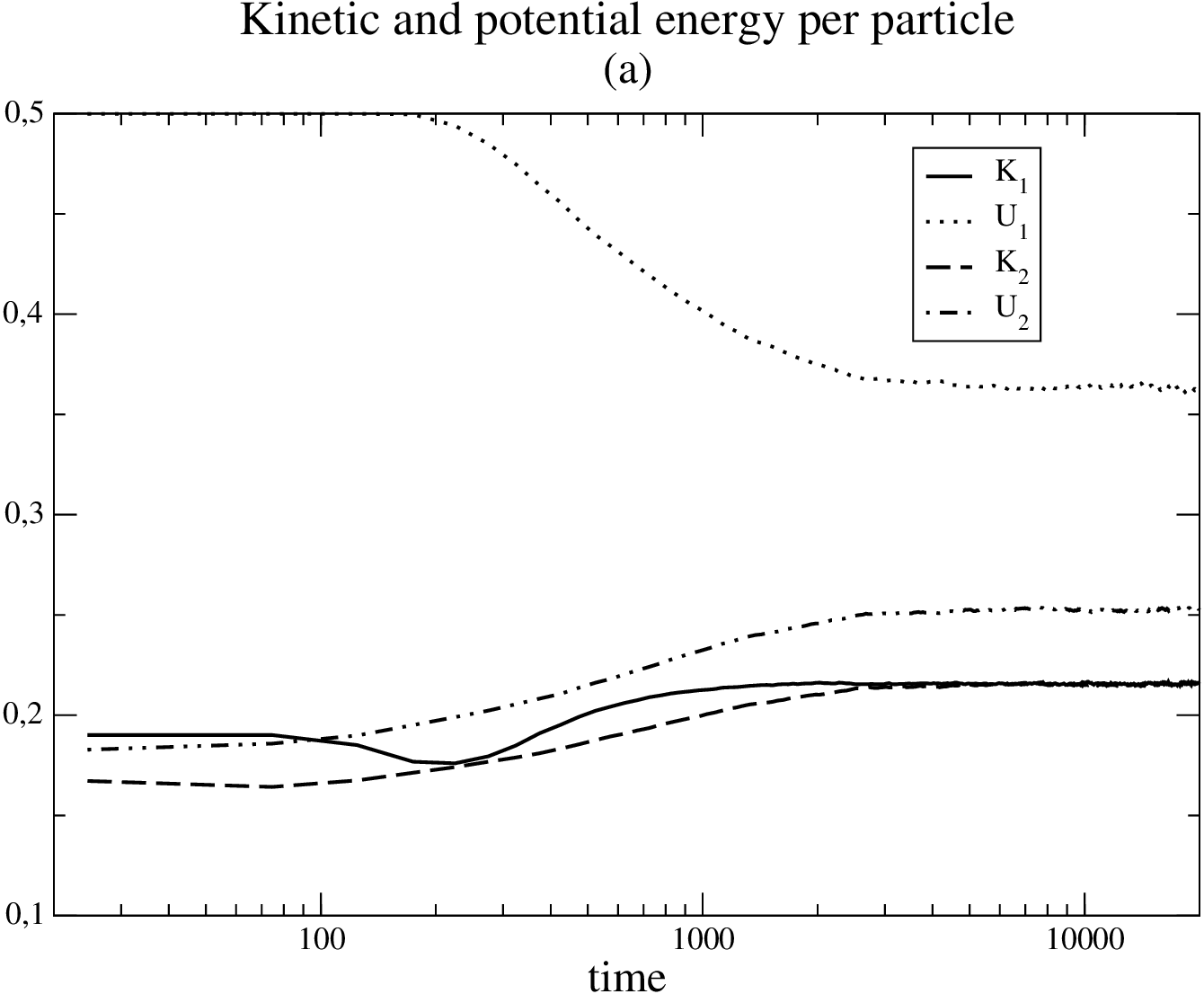}}}
\end{minipage}
\hfill
\begin{minipage}{1in}
\vspace{0mm}
\raggedright
\hspace*{-52mm}
\scalebox{0.33}{{\includegraphics{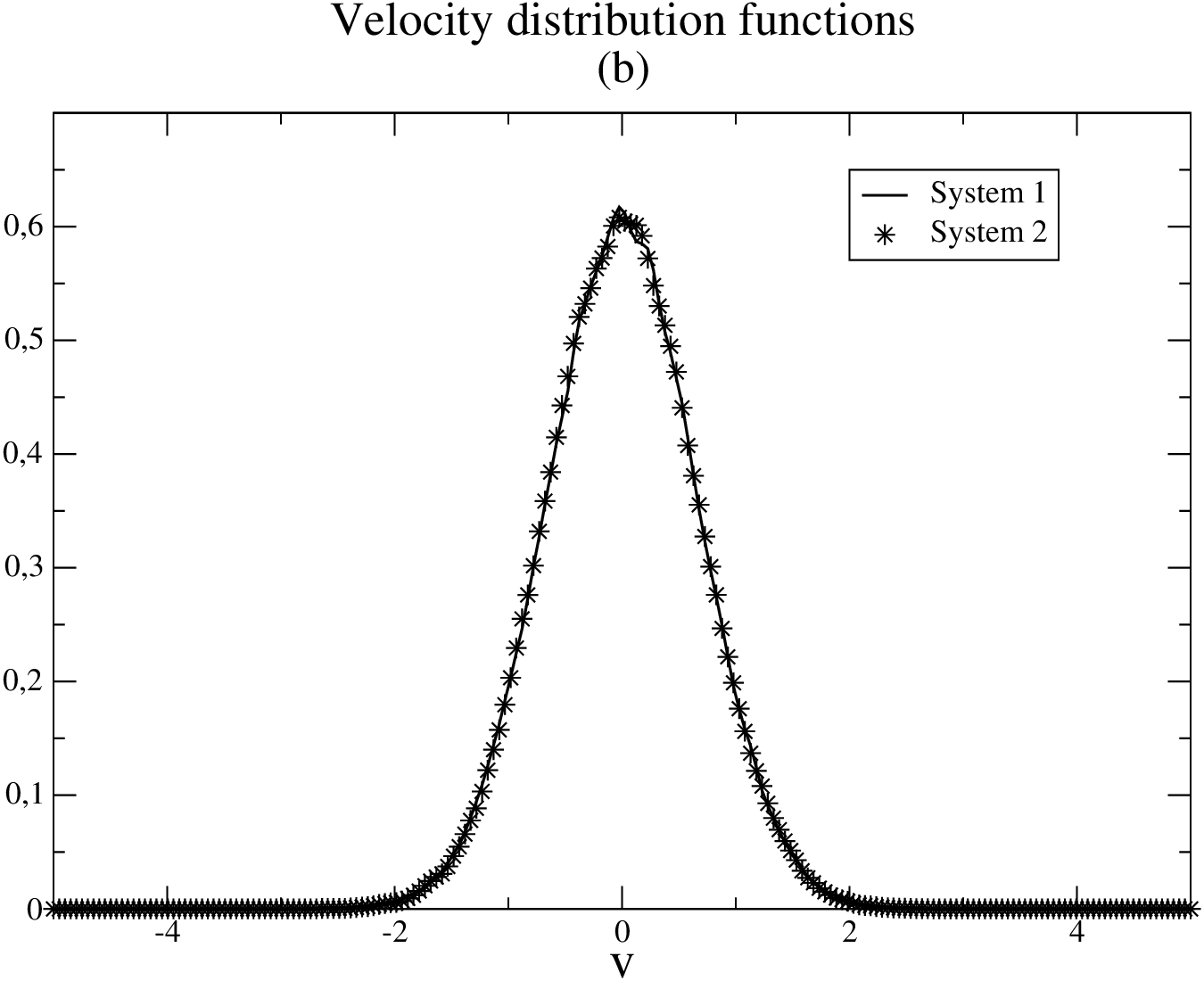}}}
\end{minipage}
\caption{(a) Kinetic and potential energies per particle with $N_1=100,000$ and $N_2=100,000$
for only one simulation and (b)
velocity distribution functions for both systems at $t=20,000$.
The interaction is turned on at $t=100$.}
\label{fig2}
\end{figure}

Once the form of the entropy is known, the temperature of the system may be defined
by $1/T=\partial S/\partial E$. This is not necessarily
identical to the Lagrange multiplier $\lambda_H$ associated to the energy constraint. The temperature $T$
is not the only relevant parameter to characterize the meta-equilibrium states of the system
corresponding to the stationary
states of the Vlasov equation. Indeed for two similar systems, with all its constituents interacting 
with the same long-range force, the condition of statistical equilibrium (most probable state)
is such that the temperature and the derivative of $S$ with respect to to
all Casimirs have the same value for both systems.
On the contrary if one tries to probe the temperature of the system using a smaller system (thermometer)
in such a way that its interaction is different in nature, i.~e.\ a short range
interaction, then the equilibrium state will be attained only after a very long time, of
the order of the relaxation time of the whole system, when both systems reach a
Maxwell-Boltzmann velocity distribution. In fact this is an expected behavior since even a small system is sufficient
to break the time invariance of the Casimirs. In order to illustrate this point, the HMF model
is modified to include a term that describes a
system with a short-range interaction ans a coupling term~\cite{r13}:
\begin{equation}
H=\sum_{i=1}^{N_1+N_2}\frac{p_i^2}{2}+\frac{1}{N_1}\sum_{i<j=1}^{N_1}v(\theta_i-\theta_j)
+\sum_{i=N_1+1}^{N_1+N_2} v(\theta_i-\theta_{i+1})
+ \sigma \sum_{i=1}^{N_2} v(\theta_i-\theta_{i+N_1}),
\label{eq1e2}
\end{equation}
where $\sigma$ is a coupling constant, $N_1$ and $N_2$ the number of particles in system 1
(with long-range interaction) and system 2 (with short-range interaction), respectively.
We integrate numerically the hamiltonian equations using the fourth order simpletic integrator of
Reference~\cite{r12c}. For system 1 we have chosen the well-studied ``water-bag'' initial condition with
an uniform distribution for the angular moments in the interval $[-p_0,p_0]$ and a completely
uniform distribution for the angles, and $p_0=\sqrt{6U_1-3}$ where $U_1$ is the energy per particle,
and the total potential energy is $N_1/2$.
This state is the limit of a family of analytic distributions of the form
$
f(\theta,p)=C\left\{1+\tanh\left[a\left(p_0^2+p^2)\right)\right]\right\},
$
for $a\rightarrow\infty$ ($C$ is a normalization constant).
The water-bag state
is stable for an energy per particle $U>7/12$~\cite{r5}.
We also consider for system 2 a water-bag initial condition but with $\theta_i=0$. In order to have
a small thermal coupling between the two systems we fix $\sigma=0.05$. It can be easily shown that the
temperature for the water-bag state is also half the kinetic energy per particle.
Figure~\ref{fig1} shows the time evolution for the kinetic and potential energies of both systems
with $N_1=100,000$ and $N_2=100$ in such a way that system~2 acts like a small thermometer. System~1
stays in the water-bag state, while system~2 do not thermalize. Eventually after a very long time
they will both reach the standard canonical equilibrium.
The attempt to define a thermometer to measure the temperature of these systems as reported by
Baldovin et al.\ \cite{r13} works only for a very special type of initial condition, and
cannot be reproduced for more general states.
In fact if the long-range interacting system is in contact with a thermal bath at a given temperature,
the only possible equilibrium is the Maxwellian distribution. This is essentially the reason why
the microcanonical and canonical ensembles are not equivalent.
Figure~\ref{fig2}a shows the kinetic and potential energies for the case $N_1=N_2=100,000$. Now
the water-bag initial condition is rapidly perturbed when the interaction is turned on and the system
evolves to a gaussian velocity distribution, as shown in Figure~\ref{fig2}b.

We have shown that the condition of maximal entropy subject to the normalization, energy
and Casimir constraints is equivalent to the non-linear stability of the stationary states
of the Vlasov equation. Therefore, in the mean-field approximation ($N\rightarrow\infty$),
the entropy for a system with long-range interaction
with a well behaved distribution function depends 
on an enumerable infinite set of parameters, i.\ e.\ the Lagrange multipliers $\lambda_I$, $\lambda_H$
and $\lambda_n$ for $n=2,\ldots,\infty$. These parameters depend in a complicated way on the values
of the constraints defined by the initial condition, the latter being  difficult to determine in realistic systems.
The proposal of the Tsallis statistics, which involves only two free parameters, the temperature
and the entropic index $q$, to describe the meta-equilibrium distribution in systems
with long-range interaction is therefore limited in scope. It can only describe a
specific type of stationary state among an infinity of different possibilities. Even for a finite
$N$ system it was shown in Reference~\cite{r5} that for a water-bag initial condition the system
eventually reaches a stationary state with an exponential tail in the distribution, contrary
to the power law behavior of the Tsallis statistics. The reasonable fitting of simulation data
obtained from the Tsallis functional form stems from the fact that it depends on two parameters,
while the Maxwellian depends only on one. Therefore for an even distribution function one can obtain
a correct fit up to its fourth moment, while only the dispersion can be fitted correctly using
a Maxwellian distribution. It is a well known fact that distribution functions that are well behaved
and that have the same first four moments are usually very close. This explains why the Tsallis
statistics can give good fitting up to some accuracy.
The small differences between the fitted function and the real function can nevertheless be essential
as for instance if one needs a correct form for the tails of the distribution.
Also the equilibrium properties of such systems are complicated to study since the type
of coupling is essential to determine which constraints are kept, while
for usual systems only the energy constraint is present and is always preserved.

The Boltzmann-Gibbs entropy is then the correct form to be used. As an importante consequence we
note that for a homogeneous state the mean field force vanishes and the
entropy is computed using eqs.~(\ref{eq6}) and $f=\Phi(e({\bf r},{\bf p}))$ with $U=0$,
and therefore it is both
additive and extensive, even despite the long range-nature of the interaction. For an
inhomogeneous state the situation is more complicated and both properties are usually lost. The use of
a non-extensive entropy by its own definition is meaningless and may lead to wrong
conclusions. The non-extensivity or non-additivity of the entropy results uniquely from the
usual definition of the Boltzmann-Gibbs entropy in the presence of correlations
among the constituents of the system, as in inhomogeneous states of long-range interacting
systems. The use of the Boltzmann-Gibbs entropy is equivalent to suppose that all microstates
compatible with the given constraints are equiprobable, which is a quite reasonable assumption. Using any
other definition of entropy introduces a non-equiprobability of states, which can result
from some ``hidden'' constraints not considered explicitly (e.\ g.\ the Casimir invariants).
If this is so, the type of non-equiprobability changes for different values of such constraints,
as well as the definition of the entropy, to take account of the change of the probabilities
of the microstates. Therefore a fixed form for a generalized entropy cannot take care of all
possible types of non-equiprobability.

The authors would like to thank Prof. A.\ Santana for carefully reading the preliminary version of this paper.
This work was partially financed by CNPq (Brazil).

\end{document}